\documentclass[3p,review]{elsarticle}

\usepackage{lineno,hyperref}
\usepackage{times}  
\usepackage{helvet} 
\usepackage{courier}  
\usepackage{graphicx} 

\usepackage{graphics, epstopdf, epsfig}
\usepackage{amstext}
\usepackage{subcaption}
\usepackage{multirow}
\usepackage{amsmath}
\usepackage{amssymb}
\usepackage{multicol}
\usepackage{rotating}
\usepackage{booktabs}
\usepackage{algorithm}
\usepackage{algpseudocode}
\usepackage{fixltx2e}
\usepackage{xcolor}
\usepackage{xspace}

\journal{Journal of \LaTeX\ Templates}

\begin{document}

\begin{frontmatter}

\title{A Unified Model for Recommendation with Selective Neighborhood Modeling}


\author[address1]{Jingwei Ma}
\ead{majingwei0824@gmail.com}

\author[address2]{Jiahui Wen}
\ead{wen\_jiahui@outlook.com}

\author[address1]{Panpan Zhang}
\ead{Zpp2567754145@gmail.com}

\author[address3]{Mingyang Zhong\corref{mycorrespondingauthor}}
\ead{my.zhong@hotmail.com}

\author[address2]{Guangda Zhang}
\ead{zhanggd\_nudt@hotmail.com}

\author[address4]{Xue Li}
\ead{lixue@neusoft.edu.cn}

\cortext[mycorrespondingauthor]{Corresponding author}

\address[address1]{Shandong Normal University, Jinan, China}
\address[address2]{National Innovative Institute of Defense Technology, Beijing, China}
\address[address3]{The University of Queensland, Brisbane, Australia}
\address[address4]{Neusoft Education Technology Group, Dalian, China}

\begin{abstract}
Neighborhood-based recommenders are a major class of Collaborative Filtering (CF) models. The intuition is to exploit neighbors with similar preferences for bridging unseen user-item pairs and alleviating data sparseness. Many existing works propose neural attention networks to aggregate neighbors and place higher weights on specific subsets of users for recommendation. However, the neighborhood information is not necessarily always informative, and the noises in the neighborhood can negatively affect the model performance. To address this issue, we propose a novel neighborhood-based recommender, where a hybrid gated network is designed to automatically separate similar neighbors from dissimilar (noisy) ones, and aggregate those similar neighbors to comprise neighborhood representations. The confidence in the neighborhood is also addressed by putting higher weights on the neighborhood representations if we are confident with the neighborhood information, and vice versa. In addition, a user-neighbor component is proposed to explicitly regularize user-neighbor proximity in the latent space. These two components are combined into a unified model to complement each other for the recommendation task. Extensive experiments on three publicly available datasets show that the proposed model consistently outperforms state-of-the-art neighborhood-based recommenders. We also study different variants of the proposed model to justify the underlying intuition of the proposed hybrid gated network and user-neighbor modeling components.
\end{abstract}


\end{frontmatter}

\section{Introduction}
With the prevalence of Internet, online services generate massive amount of data on a daily basis, and users are facing the problem of information overload when finding their interested items (e.g. books, movies). To address this problem, various recommendation techniques are proposed to quickly find interested information for the users. Model-based collaborative filtering is one of the most widely employed techniques for recommendation. However, it suffers from data sparseness \cite{luo2017inherently}, as most users usually give few ratings to the items, making it difficult to capture user preferences based mainly on the interaction data.

To mitigate data sparseness, many works have proposed to incorporate auxiliary information \cite{zhang2016collaborative} \cite{zhang2017joint} \cite{chen2017attentive} \cite{dong2017hybrid} \cite{yamasaki2017tag} such as texts, images for recommendation. Among the extra knowledge, neighborhood information has gained increasing popularity due to the fact that neighborhood-base approaches \cite{liu2018social} \cite{zhao2014leveraging} \cite{sedhain2017low} \cite{ren2017social} \cite{ma2019gated} \cite{ebesu2018collaborative} are a major class of collaborative filtering models. The intuition of the neighborhood-based approaches is that neighboring users usually share similar preferences, and those neighbors can be exploited to bridge unseen user-item pairs and mitigate data sparseness \cite{wang2016social} \cite{Tang2016Recommendations} \cite{gao2018recommendation}. Neighborhood methods capture localized semantics among users, which complements the overall global structure between users and items capitalized by latent factor models \cite{ebesu2018collaborative}. The advantages of neighborhood models and latent factor models lead to hybrid models such as SVD++ \cite{koren2008factorization} which joints neighborhood-based models and latent factor models to boost recommendation performance.

Recently, deep learning techniques such as the attention mechanism have wide applications in many research areas such as computer vision \cite{he2015delving}, question answering \cite{kumar2016ask} and machine translation \cite{bahdanau2014neural}. Previous works \cite{ebesu2018collaborative} \cite{ma2019gated} \cite{chen2019social} have demonstrated noticeable advantages by integrating the attention mechanism in neighborhood models for identifying similar users. The basic idea behinds the neural attention mechanism is to place higher weights on specific subsets of users in the neighborhood who share similar preferences, since not all neighbors are equally informative. For example, Ebesu et al. \cite{ebesu2018collaborative} propose a neural attention mechanism to learn a user-item specific neighborhood, and integrate the neighborhood component with a latent factor model for simultaneously capturing global user-item relations and local neighborhood-based structure. In \cite{chen2019social} and \cite{ma2019gated}, the authors aggregate neighbors in a weighted manner with an attention mechanism. However, there are several drawbacks with those models. First, those models simply aggregate all the neighbors, and ignore noises in the neighborhood which may negatively affect the model performance. Especially when the data is sparse and only a few neighbors are present, the aggregation of those neighbors has a great influence on the model performance. Second, they fail to model personalized neighborhood influence, and ignore the informativeness of the neighborhood given different users. The observation is that some users may highly depend on their neighbors for capturing preferences, while the others may not rely on their neighbors for recommendation. Finally, most of those works neglect the semantic proximity between the users and their neighbors, leading to an inefficient learning of the user-neighbor compatibility.

To this end, we propose a novel recommendation model (SNM for short) that seamlessly integrates selective neighborhood modeling and user-neighbor proximity preserving. One core component of SNM is a hybrid gated network that addresses neighborhood noises by identifying similar users from the neighborhood, and distilling these similar neighbors to produce a neighborhood representation for each user. Then, the hybrid gated network further approaches the neighborhood noises by pooling a user representation and his/her neighborhood representation into a unified vector for predicting the ranking score. Also, we propose to explicitly model user-neighbor similarities, and preserve user-neighbor proximity by predicting the users with their selected similar neighbors. Therefore, we can align the most informative neighbors for learning compact user representations. We integrate these two components into a unified model, so that they can be jointly learned and complementary to each other for the recommendation task.

The rationale of the proposed model is to capture personalized neighborhood influence for recommendation, as neighborhood information may have different influence on the decision-making process for different user-item pairs. With the hybrid gated network, we can control the information flow from the neighborhood based on the confidence that we have in the separation between similar and dissimilar neighbors. Therefore, we can not only select the most similar neighbors, but also model the credibility of the neighborhood information for recommendation. The contributions of this works can be concluded as follows:
\begin{itemize}
\item We propose a novel neighborhood-based recommendation model. The core component of the model is a hybrid gated network that is invulnerable to neighborhood noises when exploiting neighborhood information for recommendation. It first automatically separates similar neighbors from the dissimilar ones with a thresholding mechanism, and only aggregates those similar neighbors to produce the neighborhood representations. Then, it further filters out noises in the neighborhood by pooling representations of users and their neighborhood while considering the confidence level of the neighborhood information. Therefore, we are able to select the most informative neighbors and encode the credibility of neighborhood information for recommendation.

\item We explicitly preserve user-neighbor proximity for learning compact user representations. The user-neighbor similarities are captured by predicting users with their neighbors. Since the neighborhood representations are parameterized by the users, neighbors and the target items, user representations are learned by attending to the informative neighbors and specified for the recommendation task. We integrate the hybrid gated network and user-neighbor proximity components into a unified model, where they can mutually complement and reinforce each other to enhance the recommendation performance.

\item We validate the effectiveness of the proposed model with three publicly available datasets, and demonstrate its advantage over the state-of-the-art models. We also study different variants of the proposed model to justify the intuitions underlying each of its components.
\end{itemize}

\section{Preliminaries}
\subsection{Problem Definition}
In a recommendation problem, we have a user set $U=\{u_1,u_2,...,u_M\}$ and an item set $V=\{v_1,v_2,...,v_N\}$, where $M$ and $N$ are the number of users and items respectively. The interactions between the users and items can be denoted by a rating matrix $\mathbf{R}\in\{0,1\}^{M\times N}$, where an entry $r_{ij}=1$ in $\mathbf{R}$ means that user $u_i$ has positively rated item $v_j$. As we focus on implicit feedback recommendation in this work, the missing entries (i.e. $r_{ij}=0$) are viewed as unobserved records, and they need to be predicted. Similar to \cite{ebesu2018collaborative}, we denote $N(v_j)$ as the set of all users (neighborhood) that have provided implicit feedback for item $v_j$.


Given the rating matrix and neighborhood information, the task of this work is to jointly learn user/item representations (e.g. $\mathbf{u}_i,\mathbf{v}_j$) and predict the missing values in $\mathbf{R}$, and recommend items with high predicted values (i.e. $\hat{r}_{ij}$) for each user.

\subsection{Deep User-Item Interaction}
\label{sec:ui}
Recent works \cite{he2017neural} \cite{manotumruksa2017deep} \cite{he2018outer} employ deep neural networks for deeply modeling user-item interactions. Specifically, given the latent vectors of a user-item pair, $\mathbf{u}_i$ and $\mathbf{v}_j$ respectively, we concatenate them into a vector, and input the vector through a multi-layer perceptrons and produce a user-item interaction vector $\mathbf{z}_{ij}$, which is later for predicting the ranking score $\hat{r}_{ij}$:

\begin{equation}
\begin{split}
\mathbf{z}_{ij} &= \phi_L(...\phi_2(\phi_1(\mathbf{z}_0))...)\\
\phi_l &= \sigma_l(\mathbf{W}_l^T\mathbf{z}_{l-1}+\mathbf{b}_l),\quad l\in[1,L]\\
\mathbf{z}_0 &= [\mathbf{u}_i;\mathbf{v}_j; \mathbf{u}_i\circ \mathbf{v}_j]\\
\end{split}
\end{equation}
where $[;]$ and $\circ$ are the concatenation operation and element-wise multiply operation respectively. $\phi_l$ is the $l$-th layer neural network, and $\sigma_l, \mathbf{W}_l, \mathbf{b}_l$ are the corresponding activation function, weight matrix and bias vector respectively.

\section{The Proposed Model}
\subsection{Overview}
\begin{figure}
\centering
\includegraphics[width=7cm]{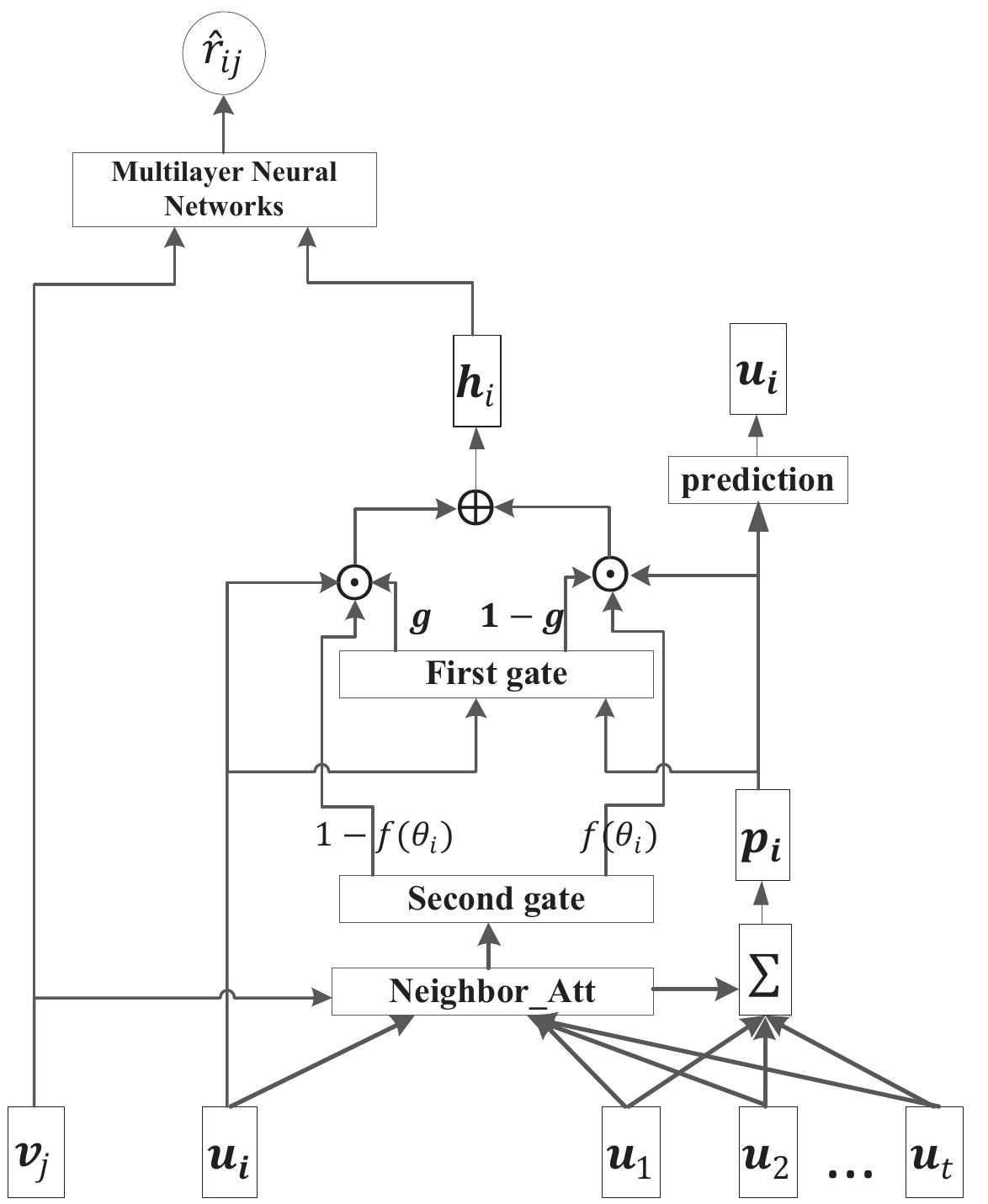}
\caption{Overview of the proposed model.}
\label{fig:overview}
\end{figure}
The overview of the proposed model for estimating a ranking score $\hat{r}_{ij}$ is illustrated in Fig.\ref{fig:overview}. As shown in the figure, given the embeddings of a user-item pair, $\mathbf{u}_i, \mathbf{v}_j$, and the corresponding embeddings of the neighbors $N(v_j)$, $\{\mathbf{u}_{1},\cdots, \mathbf{u}_{t}\}$. We first calculate the relevance scores between the user and his/her neighbors. Based on the relevance scores, we propose a hybrid gated network \cite{wang2016social} to filtering out the dissimilar neighbors and aggregate those similar neighbors into a neighborhood representation $\mathbf{p}_i$. After that, the hybrid gate network pools $\mathbf{u}_i$ and $\mathbf{p}_i$ into a unified neighborhood-based user representation $\mathbf{h}_i$. The first gate $\mathbf{g}$ is learned from the data and parameterized by $\mathbf{u}_i$ and $\mathbf{p}_i$, while the second gate $f(\theta_i)$ is based on the relevance scores and encoded domain knowledge. Finally, $\mathbf{h}_i$ and $\mathbf{v}_j$ are input to a multilayer neural network to drive the ranking score $\hat{r}_{ij}$. In addition, we propose to learn user-neighbor hierarchy for capturing compatibility between users and their neighbors. The neighborhood representation $\mathbf{p}_i$ is used for predicting the user embedding $\mathbf{u}_i$, so that the relations between users and their neighbors can be well preserved.

\subsection{Attentive Neighborhood Selection}
\label{ssec:ans}
The underlying rationale of neighborhood attention is that users with similar rating baheviors are more likely to provide similar implicit feedback to a given item. Due to the data sparsity of recommendation data, those neighbors can be exploited to gain additional insight about the existing user and item relations. As not all the neighbors are equally informative, we propose an attention mechanism to capture the most informative neighbors for neighborhood modeling.
\begin{equation}
\mathbf{p}_i = \sum_{u_t\in N(v_j)}\alpha_t \mathbf{u}_t
\end{equation}
where $\mathbf{p}_i$ is the neighborhood representation of user $u_i$, and $\alpha_t$ is the attention score assigned to one of the neighbors $u_t$ for comprising $\mathbf{p}_i$, and it is parameterized by the interactions among $u_i$, $u_t$ and $v_j$:
\begin{equation}
\label{eq:beta}
\begin{split}
&\beta_t = \mathbf{v}^Ttanh(\mathbf{W}_{ut}^T(\mathbf{u}_i\circ\mathbf{u}_t)+\mathbf{W}_{tj}^T(\mathbf{u}_t\circ\mathbf{v}_j)+\mathbf{b}_u)\\
&\alpha_t = \frac{exp(\beta_t)}{\sum_{u_t\in N(v_j)}exp(\beta_t)}
\end{split}
\end{equation}
where the matrices $\mathbf{W}_{ut}$, $\mathbf{W}_{tj}$ and the vectors $\mathbf{v}$, $\mathbf{b}_u$ are model parameters. The rationale is as follows, $\mathbf{u}_i\circ\mathbf{u}_t$ models the similarities of rating behaviors between the user $u_i$ and his/her neighbors who have rated item $v_j$, while $\mathbf{u}_t\circ\mathbf{v}_j$ captures the preferences of the neighbors over the target item $v_j$. Therefore, a neighbor yields higher relevance score $\beta_t$ if it has similar rating history as the target user $u_i$, and meanwhile has high confidence of supporting the recommendation of item $v_j$.

\paragraph{Thresholding Mechanism} Even though the attention mechanism is proposed to focus on or place higher weights on specific users in the neighborhood, simply aggregating all the neighbors can inevitably introduce noises and weaken the impact of informative neighbors. To address this problem, we employ a thresholding mechanism \cite{wang2016social} to filter out noisy neighbors, and control the information flows from the neighborhood representation to the calculation of the ranking score.

With the thresholding mechanism, the calculation of attention scores can be reformulated as follows,
\begin{equation}
\alpha_t = \frac{I(\beta_t,\theta_i)exp(\beta_t)}{\sum_{u_t\in N(v_j)} I(\beta_t, \theta_i)exp(\beta_t)}
\end{equation}
where $I(\beta_t, \theta_i)$ is an indication function, and it is used to filter out the neighbors whose relevance scores (i.e. $\beta_t$) with the target user $u_i$ are lower then a threshold $\theta_i$. The indication function is defined as follows:
\begin{equation}    I(\beta_t, \theta_i) =
 \begin{cases}
    1   &  \beta_t>\theta_i  \\
    0   &  \beta_t\leq\theta_i
 \end{cases}
\end{equation}
where the user-specific threshold $\theta_i$ is not a predefined constant in this work, but rather is left for the proposed model to learn. Therefore, a neighbor $u_t$ is not selected for calculating the neighborhood representation if the relevance score with the target user is below the threshold $\theta_i$.

\subsection{Hybrid Gate for Prediction}
\label{ssec:hgp}
The neighborhood representation can be incorporated for predicting the ranking scores, as it captures localized user-item relations \cite{ebesu2018collaborative} and complements the global user-item interactions described in Section.\ref{sec:ui}. In this work, for a given user $u_i$ and item $v_j$, we proposed a hybrid gated network to select between the user representation $\mathbf{u}_i$ and its neighborhood representation $\mathbf{p}_i$, and produce a unified neighborhood-based user representation $\mathbf{h}_i$:
\begin{equation}
\begin{split}
&\mathbf{g}=\sigma(\mathbf{W}_{g_1}^T\mathbf{u}_i+\mathbf{W}_{g_2}^T\mathbf{p}_i+\mathbf{b}_g)\\
&f(\theta_i) = \sigma((t_s-\theta_i)(\theta_i-t_d))-0.5\\
&\mathbf{h}_i = (1-f(\theta_i))*\mathbf{g}\circ\mathbf{u}_i+f(\theta_i)*(1-\mathbf{g})\circ\mathbf{p}_i\\
\end{split}
\end{equation}
where the matrices $\mathbf{W}_{g_1}, \mathbf{W}_{g_2}$, the vector $\mathbf{b}_g$ are model parameters. $\sigma(x) = \frac{1}{1+exp(-x)}$ is the sigmoid function that limits the output to the range [0,1]. The predictive vector $\mathbf{h}_i$ is a hybrid combination of the user representation $\mathbf{u}_i$ and its neighborhood representation $\mathbf{p}_i$ through two gates, $\mathbf{g}$ and $f(\theta_i)$. The first gate $\mathbf{g}$ is parameterized by $\mathbf{u}_i$ and $\mathbf{p}_i$, and it is automatically learned from the training data, while the second gate $f(\theta_i)$ encodes domain knowledge described as follows.

The basic idea behinds $f(\theta_i)$ is that it provides the degree of separation between similar neighbors and dissimilar neighbors. Specifically, $t_{s},t_{d}$ are the averages of the relevance scores of similar neighbors (i.e. neighbors with relevance scores exceed $\theta_i$) and dissimilar neighbors (i.e. neighbors with relevance scores smaller than $\theta_i$) respectively. If $t_s$ and $t_d$ are close to $\theta_i$, then $f(\theta_i)$ will be close to 0, indicating small differences between similar and dissimilar neighbors. In this case, there are high uncertainties in the neighborhood, and we have low confidence in the neighborhood preferences, hence lower weight is given to the neighborhood representation $\mathbf{p}_i$. On the contrary, $f(\theta_i)$ is close to 0.5 if $\theta_i$ provides a large degree of separation between the similar and dissimilar neighbors, then we have high confidence in the neighborhood and distribute equal weights to $\mathbf{u}_i$ and $\mathbf{p}_i$ for comprising the unified representation.

The unified representation $\mathbf{h}_i$ is then input to multilayer neural networks for estimation the ranking scores:
\begin{equation}
\begin{split}
\hat{r}_{ij} &= \phi_L(...\phi_2(\phi_1(\mathbf{z}_0))...)\\
\phi_l &= \sigma_l(\mathbf{W}_l^T\mathbf{z}_{l-1}+\mathbf{b}_l),\quad l\in[1,L]\\
\mathbf{z}_0 &= [\mathbf{h}_i;\mathbf{v}_j; \mathbf{h}_i\circ \mathbf{v}_j]\\
\end{split}
\end{equation}

The deep insight of modeling neighborhood representations is that they can bridge the semantic gap between unseen user-item pairs, and mitigate the data sparseness problem. In other words, an item can be ranked higher in the recommendation list of a user, as long as the item has been positively rated by any one of the similar neighbors.


\subsection{User-Neighbor Modeling}
\label{ssec:unm}
\begin{figure}
\centering
\includegraphics[width=6cm]{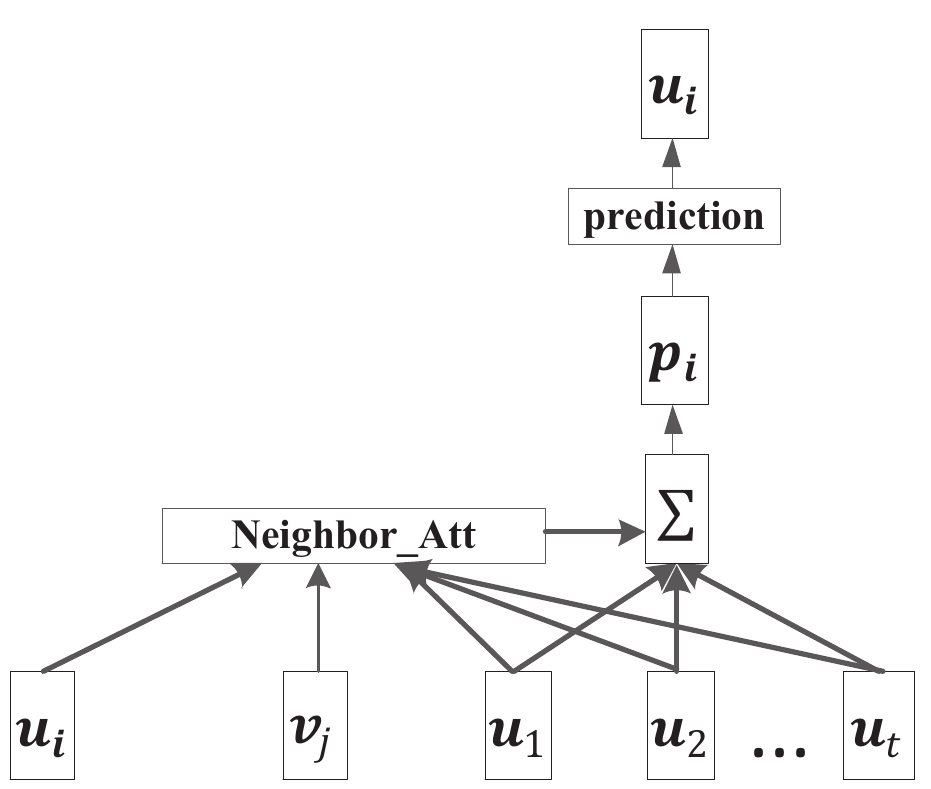}
\caption{Illustration of user-neighbor relation modeling. It is based on skip-gram model that uses a word to predict its context words or the other way around, and word embeddings are learned by maximizing the predictive probabilities. In the proposed user-neighbor modeling, users in the neighborhood are viewed as the contexts of a user, and are aggregated for predicting the target user. }
\label{fig:prediction}
\end{figure}

The intuition of modeling user-neighbor relations is to explicitly capture compatibility between users and their neighbors, as users are suppose to be close to their informative neighbors in the latent space. Notice that modeling of user-item interactions implicitly adjusts users with similar rating behaviours to have similar representations, while the proposed user-neighbor modeling explicitly captures the similarities between users and their informative neighbors. These implicit and explicit modeling of user similarities can be complementary to each other for learning compact user representations.

The proposed user-neighbor modeling is based on skip-gram model \cite{mikolov2013distributed}, which uses a word to predict its contextual words or the other way around \cite{yang2017bridging} \cite{cai2018medical}, and learns word representations by maximizing the predictive probabilities over the words. Similarly, as shown in Fig.\ref{fig:prediction}, in the proposed user-neighbor modeling, the neighborhood representation of a user can be viewed as its context. The intuition is that users with similar neighborhood representations are more likely to have similar rating behaviours, and are supposed to have similar representations in the latent space. Therefore, the user-neighbor relations are preserved by maximizing the probability of observing a user given his/her neighbors:
\begin{equation}
P(\mathbf{u}_i|\mathbf{p}_i) = \frac{exp(\mathbf{u}_i^T\mathbf{p}_i)}{\sum_{u\in U}exp(\mathbf{u}^T\mathbf{p}_i)}
\end{equation}
Notice that the neighborhood representation $\mathbf{p}_i$ is attentive summation over the neighbors, which filters out dissimilar neighbors and places higher weights on specific users in the neighborhood for a given target item $v_j$. Therefore, instead of treating the neighbors indiscriminately, the proposed user-neighbor modeling attends differently to the neighbors, finds the most informative users in the neighborhood for learning user representations. Moreover, the neighbors of a given user vary given different target items, hence user representations are learned for the specific task of recommendation.

\subsection{The Unified Model}
We integrate the hybrid gated network and user-neighbor modeling into a unified model, so that the model parameters can be jointly learned toward the optimization for the recommendation task.  In the case of implicit feedback, an entry in the rating matrix equals 1 if the item is observed and 0 otherwise. Due to the data sparseness of the recommendation problem \cite{luo2019non},  there is a large volume of unobserved items compared to the few items rated by the users. Therefore, in practice we randomly sample unobserved items as the negative items, and define binary cross-entropy loss over the estimated ranking scores and the ground truths:

\begin{equation}
\mathcal{L}_{uv} = -\sum_{(u_i,v_j)\in \mathcal{D}}(r_{ij}log(\hat{r}_{ij})+(1-r_{ij})log(1-\hat{r}_{ij}))
\end{equation}
where $\mathcal{D}$ is the training set that consists of observed user-item pairs and randomly sampled unobserved user-item pairs. As for the user-neighbor modeling, the training objective can be obtained by taking the negative log-likelihood of the conditional probability of a user given his/her neighbors:
\begin{equation}
\mathcal{L}_{u} = -\sum_{(u_i,v_j)}log\frac{exp(\mathbf{u}_i^T\mathbf{p}_i)}{\sum_{u\in U}exp(\mathbf{u}^T\mathbf{p}_i)}
\end{equation}
As shown in the equation, the calculation of $\sum_{u\in U}exp(\mathbf{u}^T\mathbf{p}_i)$ requires the summation over all users, and it incurs high computational overhead. To approach this problem, we employ negative sampling \cite{mikolov2013distributed} to approximate $log\frac{exp(\mathbf{u}_i^T\mathbf{p}_i)}{\sum_{u\in U}exp(\mathbf{u}^T\mathbf{p}_i)}$:
\begin{equation}
log\frac{exp(\mathbf{u}_i^T\mathbf{p}_i)}{\sum_{u\in U}exp(\mathbf{u}^T\mathbf{p}_i)}\approx log\sigma(\mathbf{u}_i^T\mathbf{p}_i)+\sum_{k=1}^Klog\sigma(-\mathbf{u}_k^T\mathbf{p}_i)\\
\end{equation}
where $K$ is the number of randomly sampled negative samples. Therefore, the objective function of the unified model can be defined as:
\begin{equation}
\label{eq:object}
\begin{split}
\mathcal{L} &= \mathcal{L}_{uv}+\alpha(\mathcal{L}_{u})\\
&=-\sum_{(u_i,v_j)\in \mathcal{D}}\big\{ r_{ij}log(\hat{r}_{ij})+(1-r_{ij})log(1-\hat{r}_{ij})\\
&+\alpha\big[log\sigma(\mathbf{u}_i^T\mathbf{p}_i)+\sum_{k=1}^Klog\sigma(-\mathbf{u}_k^T\mathbf{p}_i) \big]\big\}
\end{split}
\end{equation}
where $\alpha$ is the trade-off between rating score estimation and user-neighbor modeling. We optimize the objective function $\mathcal{L}$ with Adam optimizer \cite{kingma2014adam}, which is a variant of Stochastic Gradient Descent with a dynamically tuned learning rate, and updates parameters every step along the gradient direction with the following protocol:
\begin{equation}
\label{eq:learn}
\boldsymbol{\theta}^{t}\leftarrow\boldsymbol{\theta}^{t-1}-lr\frac{\partial\mathcal{L}}{\partial\boldsymbol{\theta}}
\end{equation}
where $lr$ is the learning rate, and $\boldsymbol{\theta}$ are the model parameters and $\frac{\partial\mathcal{L}}{\partial\boldsymbol{\theta}}$ are the partial derivatives of the objective function with respect to the model parameters, and they can be automatically computed with typical deep learning libraries. The overall learning algorithm of the unified model is illustrated in Algorithm.\ref{alg:model}.

\begin{algorithm}
\caption{Learning algorithm of SNM.}
\label{alg:model}
\begin{algorithmic}[1]
  \Require
  the training set $\mathcal{D}$; the learning rate $lr$; the regularization parameter $\alpha$;
  \Ensure
  latent factors of each user $u_i$ and item $v_j$, $\mathbf{u}_i$ and $\mathbf{v}_j$; semantic representation of each user $u_i$ and $v_j$; the model parameters,$\boldsymbol{\theta}$.
  \State Initialize all model parameters $\boldsymbol{\theta}$;
  \While{not convergence}
  	\State Randomly sample a tuple $(u_i,v_j)\in\mathcal{D}$;
  	\State Calculate objective loss as descried in Eqn.\ref{eq:object};
  	\State Calculate $\frac{\partial\mathcal{L}}{\partial\boldsymbol{\theta}}$ and update $\boldsymbol{\theta}$ by Eqn.\ref{eq:learn};
  \EndWhile
\end{algorithmic}
\end{algorithm}

\subsection{Time Complexity Analysis}
For each user-item pair $(u_i,v_j)$, the time complexity for selecting attentive neighbors (i.e. Section.\ref{ssec:ans}) is $O(|N(v_j)d^2|)$, where $d$ is the embedding size and $|N(v_j)|$ is the number of concerned neighbors. The time complexity for hybrid gate prediction (i.e. Section.\ref{ssec:hgp}) is $O(d^2+\sum_{l=1}^Ld_{l-1}d_l)$, where $O(d^2)$ is the time complexity for computing the predictive vector $\mathbf{h}_i$, $O(\sum_{l=1}^Ld_{l-1}d_l)$ is the time complexity of the multilayer neural networks and $d_l$ is the dimension of the $l$-th layer network. As for the user-neighbor modeling (i.e. Section.\ref{ssec:unm}), the time complexity is $O((K+1)d)$ where $K$ is the number of negative neighbors. Therefore, the time complexity of the proposed model is $O(|N(v_j)|d^2+d^2+\sum_{l=1}^Ld_{l-1}d_l+(K+1)d)$. In practice, $|N(v_j)|,K$ are much smaller than $d$, hence the time complexity of SNM mainly depends on the quantity of latent dimensions.



\section{Experiment}

\subsection{Datasets}
\begin{table}
\centering
\caption{Statistics of the datasets for experiment}
\begin{tabular}{l|llll}
\toprule
\textbf{Dataset} & \textbf{\#user} & \textbf{\#item} & \textbf{\#rating} & \textbf{sparsity}\\
\hline
\textit{movielen-20m} & 138,493 & 18,307 & 19,977,049 & 99.22\% \\
\textit{Pinterest} & 55,187 & 9,916 & 1,500,809 & 99.73\%\\
\textit{citeulike-a} & 5,551 & 16,980 & 204,987 & 99.78\%\\
\bottomrule
\end{tabular}
\label{tb:data}
\end{table}
In this section, we validate the effectiveness of the proposed model with three publicly available datasets. The first dataset is \textit{movieLens-20m} \footnotemark[1] \cite{ma2019gated} where users provide explicit ratings and reviews toward movies. To transform the explicit ratings into implicit feedbacks, ratings higher than 3 are regarded as positive feedbacks, and the others are regarded as unobserved interactions. The second dataset is \textit{Pinterest} \cite{geng2015learning} where users can save or pin an image that they are interested in. A positive feedback is recorded if a user save or pin an image. The third datasets \textit{citeulike-a} \footnotemark[2] \cite{wang2011collaborative} is collected from an online service that allows users to save and share academic papers. A user-item interaction is encoded as 1 if the user has saved the paper in his/her library. Similar to previous works \cite{Liang2016modeling} \cite{Lian2014geomf} \cite{yuan2013time}, we filter out users with fewer than 5 positive items and the items with fewer than 2 users. The statistics of the datasets is shown in Table.\ref{tb:data}.
\footnotetext[1]{https://grouplens.org/datasets/movielens/}
\footnotetext[2]{http://www.cs.cmu.edu/~chongw/data/citeulike/}
\footnotetext[3]{http://sites.google.com/site/xueatalphabeta/}

\subsection{Baselines}
The baselines employed for performance comparison are list as follows,
\begin{itemize}
\item SLIM \cite{ning2011slim} generates top-N recommendations by aggregating from user purchase/rating profiles.

\item NeuMF\cite{he2017neural}  combines generalized Matrix Factorization (MF) and Multi-Layer Perception (MLP) for modeling user-item latent structures.

\item SVD++\cite{koren2008factorization} is a hybrid model that encodes latent factor model and neighborhood similarity into a unified framework for recommendation.


\item GATE \cite{ma2019gated} exploits neighboring relations to help infer users' preferences.

\item SAMN \cite{chen2019social} models aspect and friend-level influences in an hierarchical manner.

\item CMN \cite{ebesu2018collaborative} identifies similar neighboring users with an attention mechanism based on the specific user-item pair, and jointly exploits the neighborhood state and user-item interactions to derive recommendation.

\item DELF\cite{cheng2018delf} proposes an attention mechanism to aggregate an additional embedding for each user/item, and then further introduce a neural network architecture to incorporate dual embeddings for recommendation.
\end{itemize}

The baselines are selected for the consideration that most of them exploit neighborhood information to alleviate the problem of data sparsity, hence they are comparable to the proposed model. The difference among those models is the way that they leverage the neighborhood compatibility. Specifically, SVD++ equally aggregates neighbors and hybrid it with the latent factor model. CMN proposes an attention mechanism to place higher weights on users in the neighborhood given the specific target item. DEFL also employs an attention mechanism to aggregate neighbors on both user and item side. For fair comparison, some baselines are modified  in this paper. For example, the content-ware component in GATE is excluded, as textual information is not available in our case. As for SAMN, user-user friendships are created based on common ratings.

\subsection{Implementation}
We implement the proposed model based on tensorflow deep-learning library \footnotemark[4]. As for the hyper-parameters, we perform a grid search for latent factors amongst \{16,32,64,128\}, and the trade-off weight between the ranking score estimation and user-neighbor modeling from \{0.001,0.01,0.1,1\}. The number of layers for modeling the user-item interactions is varied from \{1,2,3\}, with the dimensionality of each layer being halved from the previous layer. As the initialization of the deep neural network has a crucial impact on the recommendation performance, we employ the NeuMF model to pre-train the user/item embeddings, and then use them to initialize the corresponding parameters when training the proposed model. The implemented model is trained via stochastic gradient descent over shuffled mini-batches with a batch size of 256. The defined training objective is optimized using Adam \cite{kingma2014adam} optimizer with an initial learning rate of 0.001, and it is decayed with a rate of 0.9 for every 100 steps. We perform early stopping and fine tune the parameters with the dev set. All of the experiments and training are done using a NVIDIA GeForce GTX 1070 graphics card with 8G memory.

\footnotetext[4]{https://tensorflow.google.cn/api\_docs/python}

\subsection{Evaluation}
To evaluate the proposed model, we randomly split the datasets into training set (70\%), validation set (10\%) and testing set (20\%). We repeat the experiments for 10 times to avoid splitting bias, and report the average results over the 10 runs. However, we notice very minimal differences among the performances of different runs. All the competitive models are fine-tuned on the validation set. In the training process, for each positive item, we randomly sample 5 items as the negative samples. In the testing process, as it is time-consuming to rank all the items for every user at each time, hence for each $(u_i,v_j)$ pair in the testing set,  we mix the testing item with 99 random items, and rank the testing item along with the 99 items for the related user. We measure the recommendation performance with the commonly used Hit Ratio (HR) and Normalized Discounted Cumulative Gain (NDCG), as shown follows,
\begin{equation}
\begin{split}
HR&=\frac{\#hits}{\#test}\\
NDCG&=\frac{1}{\#test}\sum_{i=1}^{\#test}\frac{1}{log_2(p_i+1)}
\end{split}
\end{equation}
where $\#hits$ is the number of testing item that appears in the the recommendation list of the related user and $\#test$ is the total number of $(u_i,v_j)$ pair in the testing set. $p_i$ is the position of the testing item in the recommendation list for the $i$-th hit. HR measures whether the testing item is in the recommendation list, while NDCG assigns higher score to the testing item with higher position. In this paper, we truncate the ranking list at $k\in[1,2,\cdots,10]$ for both metrics. For example, HR@5 measures the ratio of the testing items that appears in the Top-5 recommendation list.

\subsection{Baseline Comparison}
\begin{table}
\centering
\caption{Performance comparison on three datasets. Best performance is in boldface and the second best is underlined. * indicates the results are significant at level 0.01. }
\begin{tabular}{l|cccc}
\toprule
\multicolumn{5}{c}{\textit{movielen-20m}}\\
\hline
Models&HR@5&HR@10&NDCG@5&NDCG@10 \\
\hline
SLIM&0.4021&0.4754&0.3277&0.3631 \\
NeuMF&0.4198&0.5311&0.3157&0.3515 \\
SVD++&0.4086&0.5186&0.3004&0.3356 \\
GATE&0.4523&0.5448&\underline{0.3591}&\underline{0.3922} \\
SAMN&0.4412&0.5217&0.3542& 0.3873 \\
CMN&\underline{0.4680}&0.5687&0.3509&0.3832 \\
DELF&0.4517&\underline{0.5499}&0.3462&0.3779 \\
SNM&\textbf{0.4986}*&\textbf{0.5981}*&\textbf{0.3837}*&\textbf{0.4160}*\\
\hline
\toprule
\multicolumn{5}{c}{\textit{Pinterest}}\\
\hline
Models&HR@5&HR@10&NDCG@5&NDCG@10 \\
\hline
SLIM&0.5115&0.6232&0.3797&0.4125 \\
NeuMF&0.5206&0.6371&0.3819&0.4197 \\
SVD++&0.4804&0.6075&0.3430&0.3842 \\
GATE&0.5550&0.6500&0.4262&0.4580 \\
SAMN&0.5637&0.6412&0.4243&0.4565 \\
CMN&0.5454&0.6415&0.4170&0.4482 \\
DELF&\underline{0.5699}&\underline{0.6578}&\underline{0.4283}&\underline{0.4618} \\
SNM&\textbf{0.5882}*&\textbf{0.6787}*&\textbf{0.4480}*&\textbf{0.4762}*\\
\hline
\toprule
\multicolumn{5}{c}{\textit{citeulike-a}}\\
\hline
Models&HR@5&HR@10&NDCG@5&NDCG@10 \\
\hline
SLIM&0.3932&0.5637&0.3018&0.3651 \\
NeuMF&0.4230&0.5807&0.3395&0.3997 \\
SVD++&0.4971&0.6094&0.3689&0.4017 \\
GATE&0.4995&0.6134&0.3702&0.4079 \\
SAMN&0.5094&0.6154&0.3710&\underline{0.4109}\\
CMN&0.4377&0.5996&0.3481&0.4099 \\
DELF&\underline{0.5095}&\underline{0.6177}&\underline{0.3785}&0.4090 \\
SNM&\textbf{0.5224}*&\textbf{0.6305}*&\textbf{0.3866}*&\textbf{0.4216}* \\
\bottomrule
\end{tabular}
\label{tb:cmp}
\end{table}

The performance comparison among the models is presented in Table.\ref{tb:cmp}. From the table, we have the following key observations. First, the proposed model, SNM, outperforms the baselines significantly, demonstrating the successful integration of the hybrid gated neighborhood selection and user-neighbor modeling over existing attentive neighborhood modeling methods. Second, SNM yields better results than the dual attention model, DELF. Although DELF dually exploits user and item neighboring information for recommendation, it simply incorporates aggregated neighborhood for interactions, which dose not consider the usefulness of the neighborhood information, leading to the incomplete exploration of neighborhood credibility. Third, SNM outperforms CMN, SAMN and GATE by a large margin. Those three models all aggregate neighbors into a unified vector with an attention mechanism, and incorporate it with different neural networks. However, those models do not discriminate the effects of different neighborhood, which may inevitably introduce noises and negatively affect model performance. Forth, SNM achieves better results than SVD++, since SVD++ treats neighbors equally and neglects the effects that informative neighbors can better bridge gaps between unseen user-item pairs. Finally, SNM demonstrates improved performance over NeuMF and SLIM, as those two models dose not incorporate neighborhood information. Therefore, they mainly depend on user-item interaction data and suffer from the problem of data sparseness.

\paragraph{Other observations}First, all the attention-based models (i.e. DELF, CMN, SAMN, GATE) achieve competing recommendation performance, since they all involve an attention mechanism to incorporate neighborhood information for recommendation. Second, among the attention-based models, DELF achieves overall better performance than CMN, SAMN and GATE. One possible explanation is that DELF simultaneously explores user-user and item-item relations for modeling dual interactions, while the other three models mainly depend on singular neighborhood for recommendation. Third, all the attention-based models achieve better performance than SVD++, demonstrating the benefit of placing higher weights on informative neighbors that can better infer users' preferences. Forth, SLIM generally performs worst among the baselines, since it mainly depends on user-item interactions and suffers from data sparsity. Fifth, on \textit{Pinterest} dataset, SVD++ obtains the lowest HR and NDCG, due to the restrictive ability to handle sparse data when only few neighbors are present. Finally, on dataset \textit{movielen-20m} and \textit{Pinterest}, NeuMF outperforms the neighborhood-based SVD++ revealing the effectiveness of multi-layer non-linear transformations for capturing complex user-item interactions.

\subsection{Components Study}


\begin{table}
\centering
\caption{Performance comparisons among different variants of SNM. SNM-Th excludes the hybrid gated network, where the neighborhood representation is a weighted sum over all the neighbors, and it is weighted equally as the user representation for comparison neighborhood-based user representation $\mathbf{h}_i$. SNM-Un excludes the user-neighbor modeling, and users are not regularized to be close to their neighbors. }
\begin{tabular}{p{1.8cm}|l|ccc}
\toprule
Datasets& Metrics &SNM-Th&SNM-Un&SNM\\
\hline
\multirow{4}*{\textit{movielen-20m}}& HR@5 & 0.4686&0.4828&0.4986\\
& HR@10   &0.5766&0.5789&0.5981\\
& NDCG@5   &0.3567&0.3635&0.3837\\
&NDCG@10   &0.3884&0.3929&0.4160\\

\hline
\multirow{4}*{\textit{Pinterest}} & HR@5 & 0.5702   & 0.5845& 0.5882\\
& HR@10   & 0.6653  & 0.6749 & 0.6787\\
& NDCG@5   & 0.4109 & 0.4452 & 0.4480\\
&NDCG@10   & 0.4417 & 0.4759 & 0.4762\\
\hline

\multirow{4}*{\textit{citeulike-a}} & HR@5   & 0.5191 & 0.5084 & 0.5224\\
& HR@10   & 0.6228  & 0.6112 & 0.6305\\
& NDCG@5   & 0.3866 & 0.3850 & 0.3927\\
&NDCG@10  & 0.4216 & 0.4183 & 0.4263\\
\bottomrule
\end{tabular}
\label{tab:variants}
\end{table}

In this section, we further investigate the impact that different components have on the recommendation performance. To this end, we introduce two variants of SNM, namely (a) the variant that excludes the \textbf{Th}resholding mechanism in the hybrid gated network (SNM-Th for short) and (b) the variant that excludes the \textbf{U}ser-\textbf{n}eighbor modeling (SNM-Un for short). For SNM-Th, the neighborhood representation is an attentive aggregation over the neighbors without filtering out the dissimilar users, and we place equal weights on the user representation and its neighborhood representations to comprise $\mathbf{h}_i$.

As shown in Table.\ref{tab:variants},  the unified model SNM generally outperforms SNM-Th that excludes the hybrid gated network and attends to all the neighbors for comprising the neighborhood representation. Further illustrating effectiveness of the hybrid gated network, as it can not only filter out noisy neighbors for a given item, but also capture the uncertainty in the neighborhood. SNM-Un uniformly performs worse than SNM hinting at the effectiveness of the user-neighbor modeling that explicitly captures the proximity between each user and its neighborhood in the latent space. The performance differences between SNM-Th and SNM-Un vary across the datasets. Specifically, SNM-Un yields better performance than SNM-Th on \textit{movielen-20m} and \textit{Pinterest} datasets, while SNM-Th shows performance improvement over SNM-Un on \textit{citeulike-a} dataset. This can be explained by the fact that items of \textit{movielen-20m} and \textit{Pinterest} datasets have more ratings on average, hence the denser neighborhood allows SNM-Un to sufficiently explore the neighborhood and identify the most informative neighbors for bridging unseen user-item pairs.
%
%
%

\subsection{Hyperparameter Study}
\begin{figure*}
  \centering
  \subcaptionbox{movielen-20m}[.32\textwidth][c]{%
    \includegraphics[width=.35\textwidth]{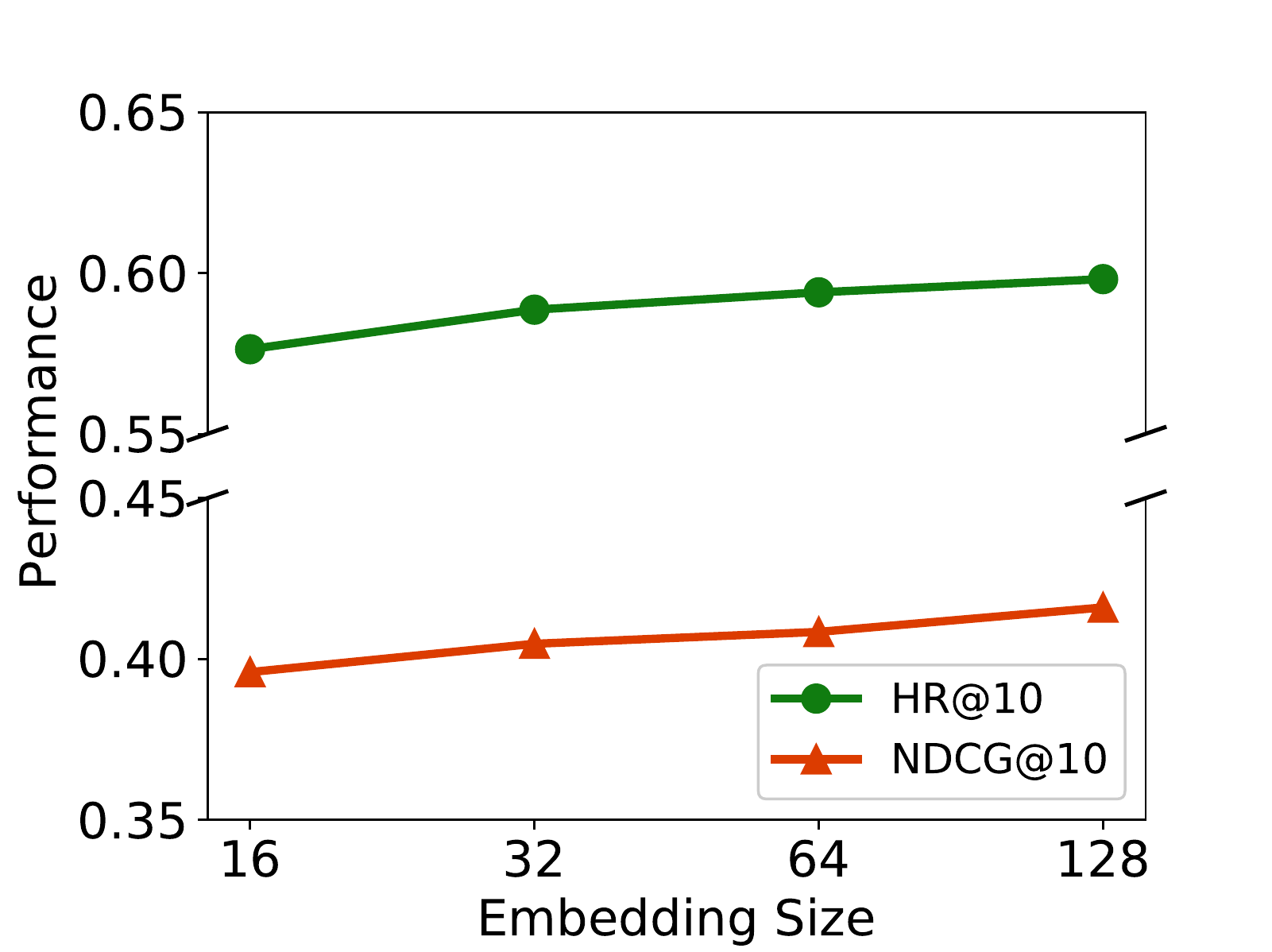}}
  \subcaptionbox{Pinterest}[.32\textwidth][c]{%
    \includegraphics[width=.35\textwidth]{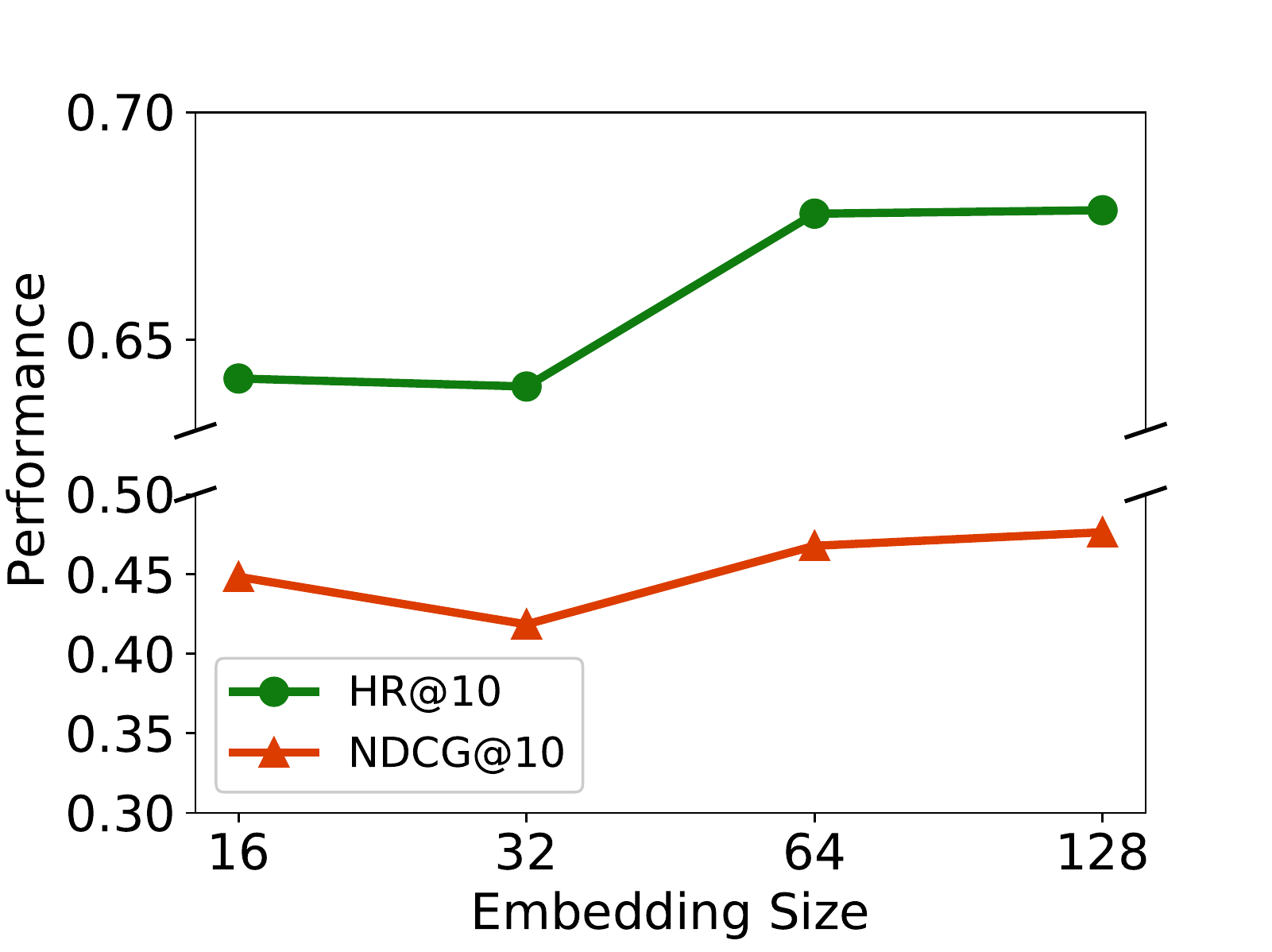}}
  \subcaptionbox{citeulike-a}[.32\textwidth][c]{%
    \includegraphics[width=.35\textwidth]{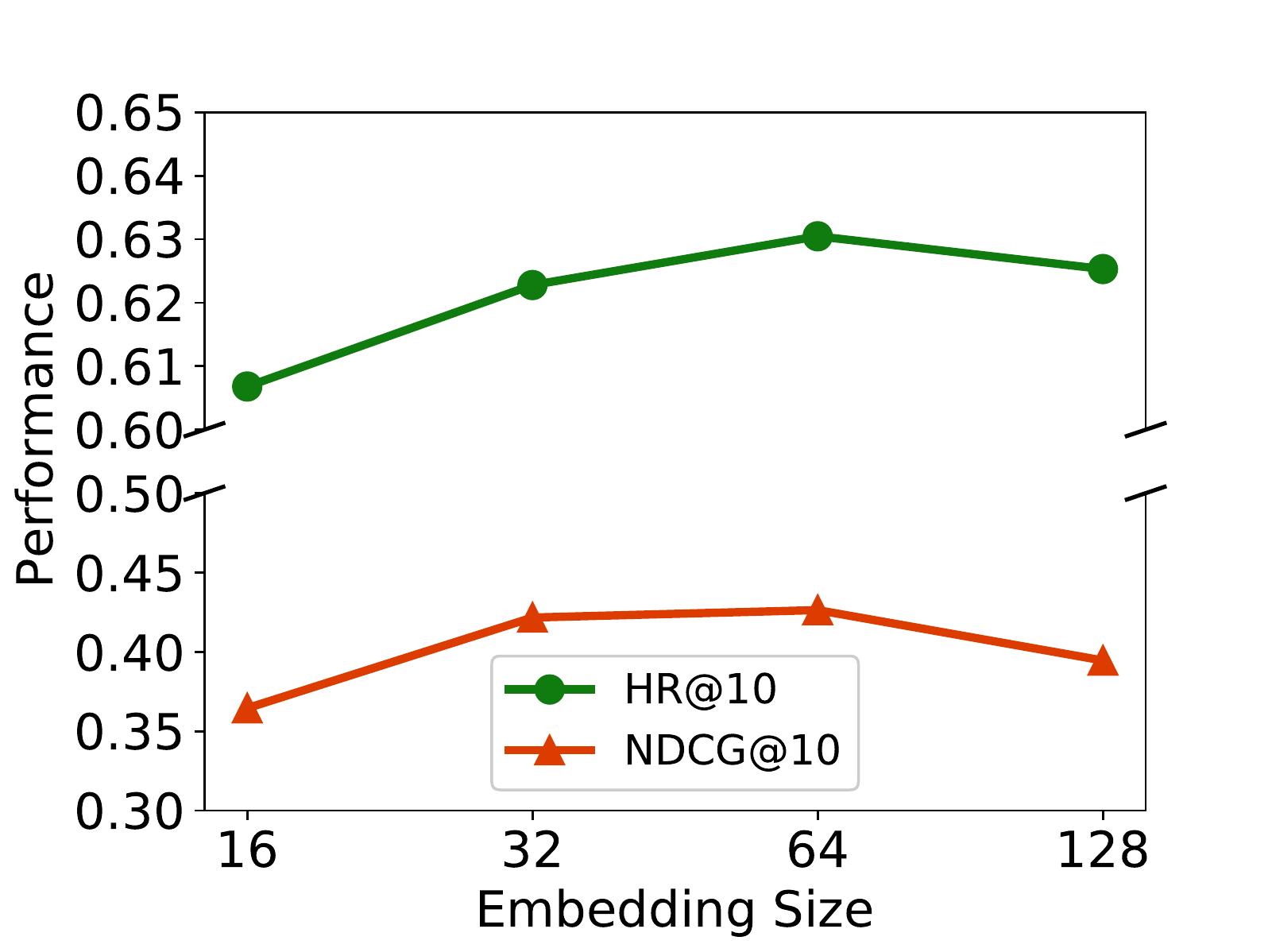}}
  \caption{Recommendation performance with respect to different embedding sizes on the three datasets. }
  \label{fig:embed}
\end{figure*}

\begin{figure*}
  \centering
  \subcaptionbox{movielen-20m}[.32\textwidth][c]{%
    \includegraphics[width=.35\textwidth]{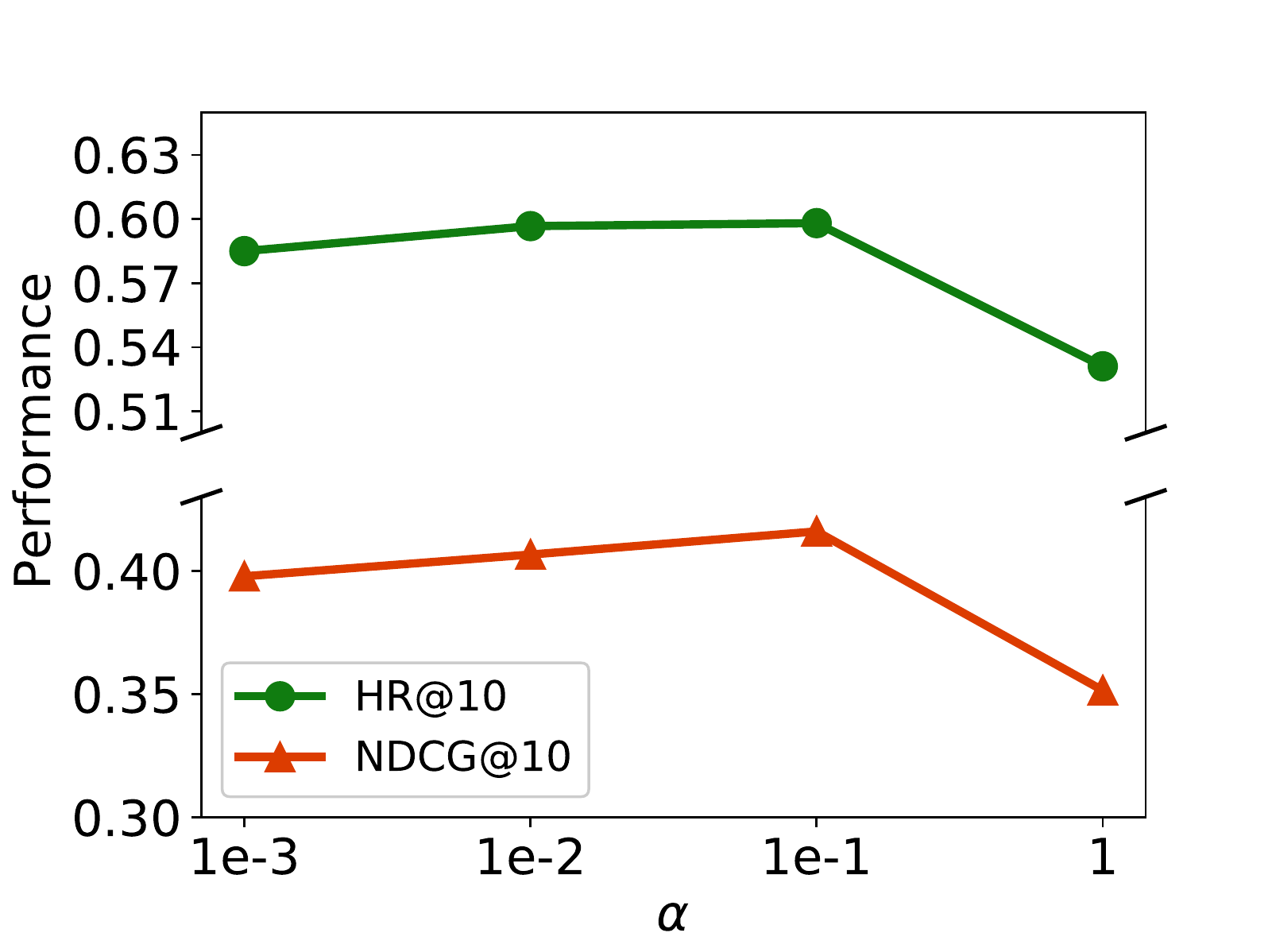}}
  \subcaptionbox{Pinterest}[.32\textwidth][c]{%
    \includegraphics[width=.35\textwidth]{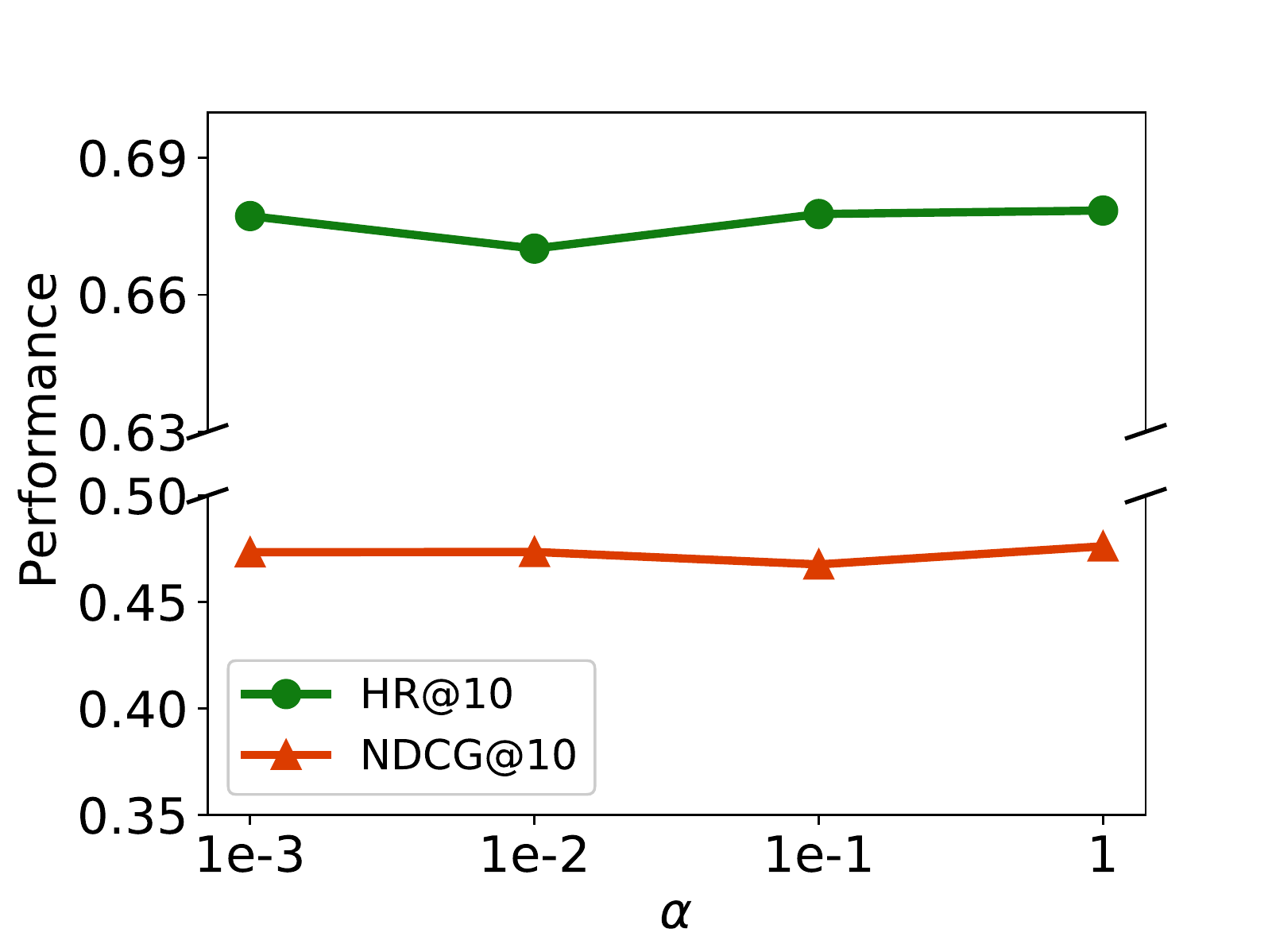}}
  \subcaptionbox{citeulike-a}[.32\textwidth][c]{%
    \includegraphics[width=.35\textwidth]{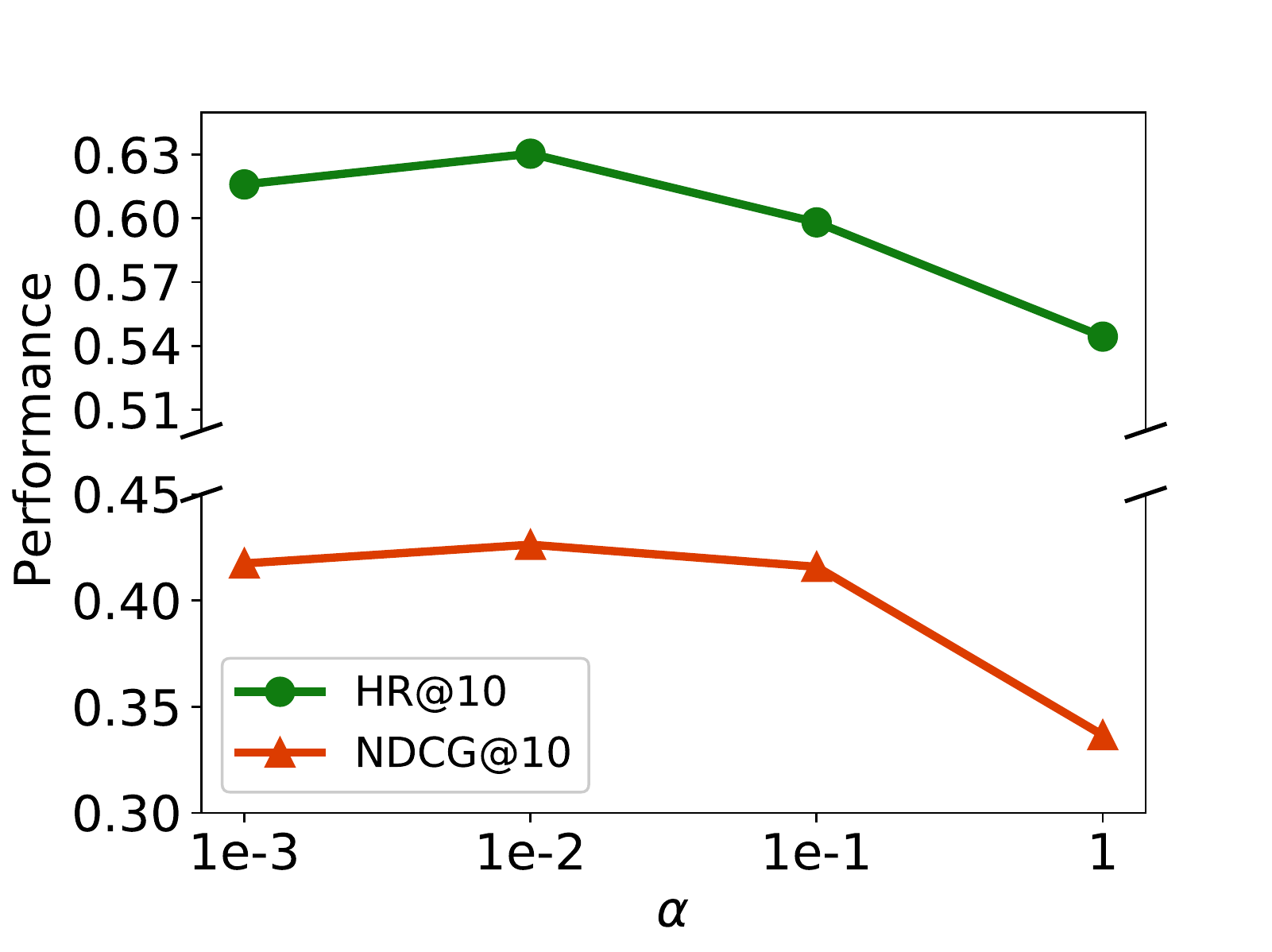}}
  \caption{Recommendation performance with respect to different regularization weights (i.e. $\alpha$ in Eqn.\ref{eq:object}) on the three datasets. }
  \label{fig:alpha}
\end{figure*}

\begin{figure}
\centering
\includegraphics[width=7cm]{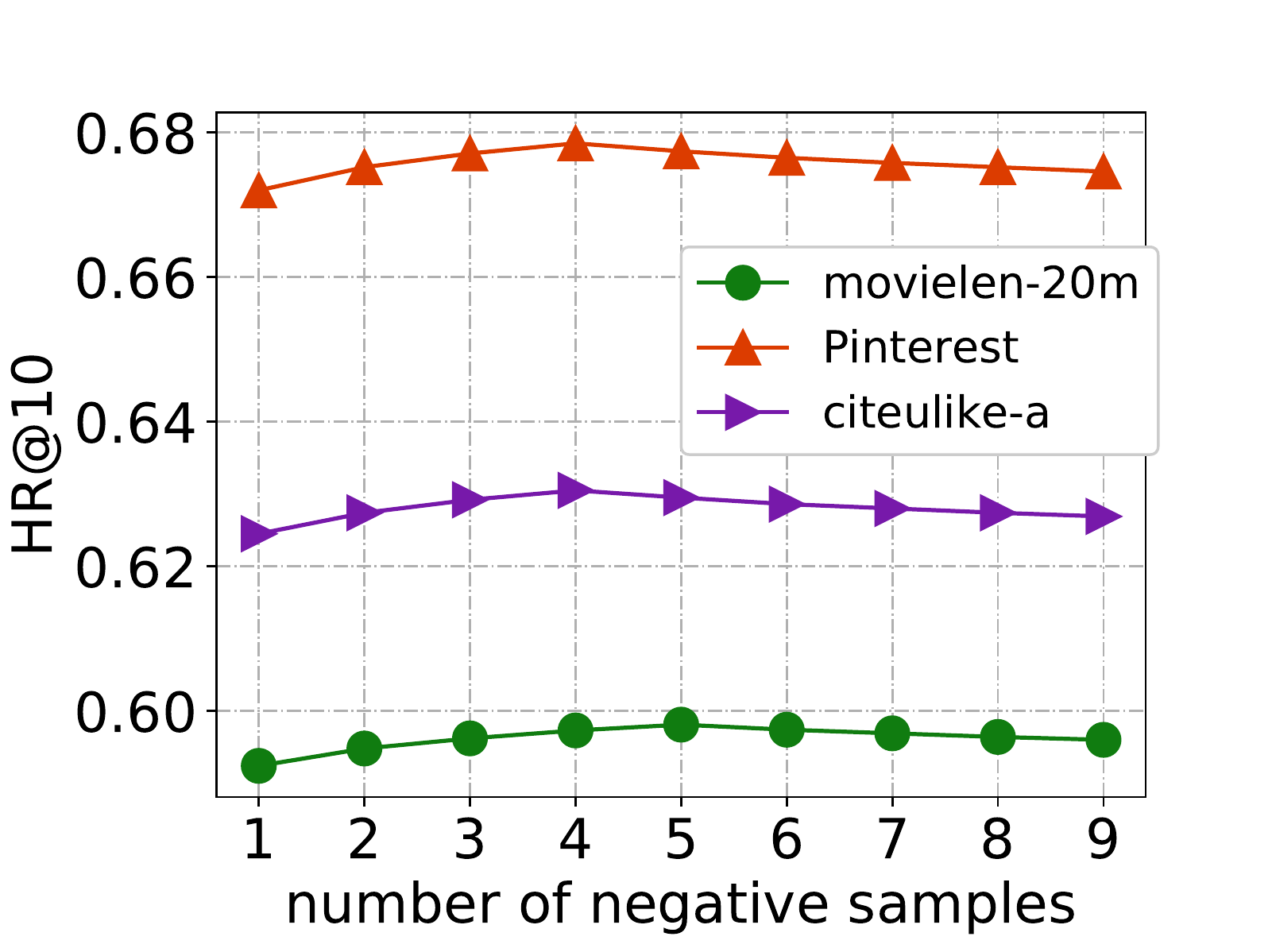}
\caption{HR@10 of SNM with respect to different ratios of negative items.}
\label{fig:negative}
\end{figure}

\paragraph{Embedding Size} The recommendation performance (i.e. HR@10, NDCG@10) of SNM with respect to different embedding sizes is presented in figure \ref{fig:embed}. As HR@10 and NDCG@10 shows similar trends, we focus the analysis on HR@10 in this section. For the \textit{movielen-20m} dataset, the general trends shows a steady improvement as the embedding size increases, indicating large embedding size improves the model's expressiveness to encode complex user-item interactions of the training data. For the \textit{Pinterest} dataset, the similar trend of model performance with the increasing embedding size can also be observed. An exception can be found when the embedding size is set to 32, where the model experiences an unusual drop in performance. A possible explanation is that the model falls into a local optimum because of nonconvexity of neural networks. Further, increasing the embedding size from 64 to 128 does not provide significant benefits, indicating the embedding size can be appropriately tuned to yield a trade off between computational overhead and model performance. Experiment results on \textit{citeulike-a} dataset show similar performance gains as the embedding size increases. However, unlike the results from previous datasets that the model achieves the best performance with the embedding size 128, the \textit{citeulike-a} dataset shows a sudden drop with that embedding size potentially due to overfitting.

\paragraph{Trade-off Weight} In this subsection, we investigate the impact that the trade-off weight (i.e. $\alpha$ in Eqn.\ref{eq:object}) has on the model performance. The trade-off weight is used to introduce the user-neighbor modeling, and regularize how the users should be close to their neighbors. Figure \ref{fig:alpha} shows the mode performance with respect to different trade-off weights across the datasets. For the \textit{movielen-20m} dataset,
the model performance shows a gradual improvement when the weight is increased from 0.001 to 0.1, signifying the benefits of explicitly capturing the proximity between the users and their neighbors. However, the model performance degrades significantly when the weight is further increased indicating strong regularizations on the users can confine the model's expressiveness to infer complex user-item relations. On the contrary, on the \textit{Pinterest} dataset, the proposed model presents constantly stable performance with respect to different trade-off weights. This is potentially due to the dense neighborhood of that dataset, hence sufficient neighborhood information makes the model invulnerable to the regularizations between users. The model performance on \textit{citeulike-a} shows similar trends as that on \textit{movielen-20m} datasets, however, the model is more sensitive to the hyperparameter. This is probably because the modeling of user-neighbor similarities causes big variances to user representations when only a few neighbors are present.

\paragraph{Negative Sampling Ratio}In this subsection, we study the influence of negative sampling ratio. Negative items are dominant in the training set, and they usually contain rich information for recommendation \cite{wu2019npa}. Figure \ref{fig:negative} shows the HR@10 of the proposed model with respect to different negative sampling ratios. From the figure, we can observe that the model performance constantly improves with increasing negative ratio. This is probably because when the ratio is small, the informativeness of the negative items can not be sufficiently exploited for boosting recommendation. However, when the ratio is too large, the model performance begins to degrade. One possible explanation is that with large negative ratio, the training set is dominated with negative items, and the model is biased to those items, which lead to sub-optimal results.

\subsection{Time Complexity Analysis}

\begin{table*}
\centering
\caption{Running time in seconds of the competing models across the datasets. The training time is the time for training an epoch of the data. }
\begin{tabular}{l|cc|cc|cc}
\toprule
\multirow{2}*{Models} & \multicolumn{2}{c|}{\textit{movielen-20m}} & \multicolumn{2}{c|}{\textit{Pinterest}} & \multicolumn{2}{c}{\textit{citeulike-a}}\\
& train & test & train & test & train & test \\
\hline
GATE & 1006.1691&50.2672&63.748&3.2282&22.3168&1.1196 \\
CMN & 1328.8509&49.7335&86.7495&3.3231&30.7609&1.1525 \\
SNM & 1215.6646&61.0406&77.2744&3.8188&27.1653&1.2702 \\
SAMN & 1538.63&69.7845&98.7012&4.4937&34.772&1.535 \\
DELF & 2650.0845&143.348&167.1174&9.2574&60.343&3.1212 \\
\bottomrule
\end{tabular}
\label{tb:time}
\end{table*}

To study the time efficiency of the proposed model, we compare its running time with some representative baselines. We record the running time across the models on the same computing environment, and set the hyperparameters according to the original works. Table.\ref{tb:time} shows the training and prediction time of the neighborhood-based models. The training time is the time for training one epoch of the data, and the prediction time is the time required to complete the prediction for the whole testing set. From the table we can observe that GATE takes the least time to finish the training and testing processes, as the component for calculating textual representations is excluded in this work. CMN and SNM yield competing results, since the time complexity mainly depends on the embedding size in both models. SAMN and DELF incur high computational overhead, as they involve hierarchical or dual attentions. For example, SAMN proposes to capture aspect attentions and friend-level attentions for user modeling, while DELF simultaneously model user-user and item-item relations for multilevel interactions. This efficiency study shows the proposed SNM model is able to achieve the best recommendation accuracy without incurring noticeable overhead.

\subsection{Visualization}
\begin{figure*}
  \centering
  \subcaptionbox{movielen-20m}[.32\textwidth][c]{%
    \includegraphics[width=.32\textwidth]{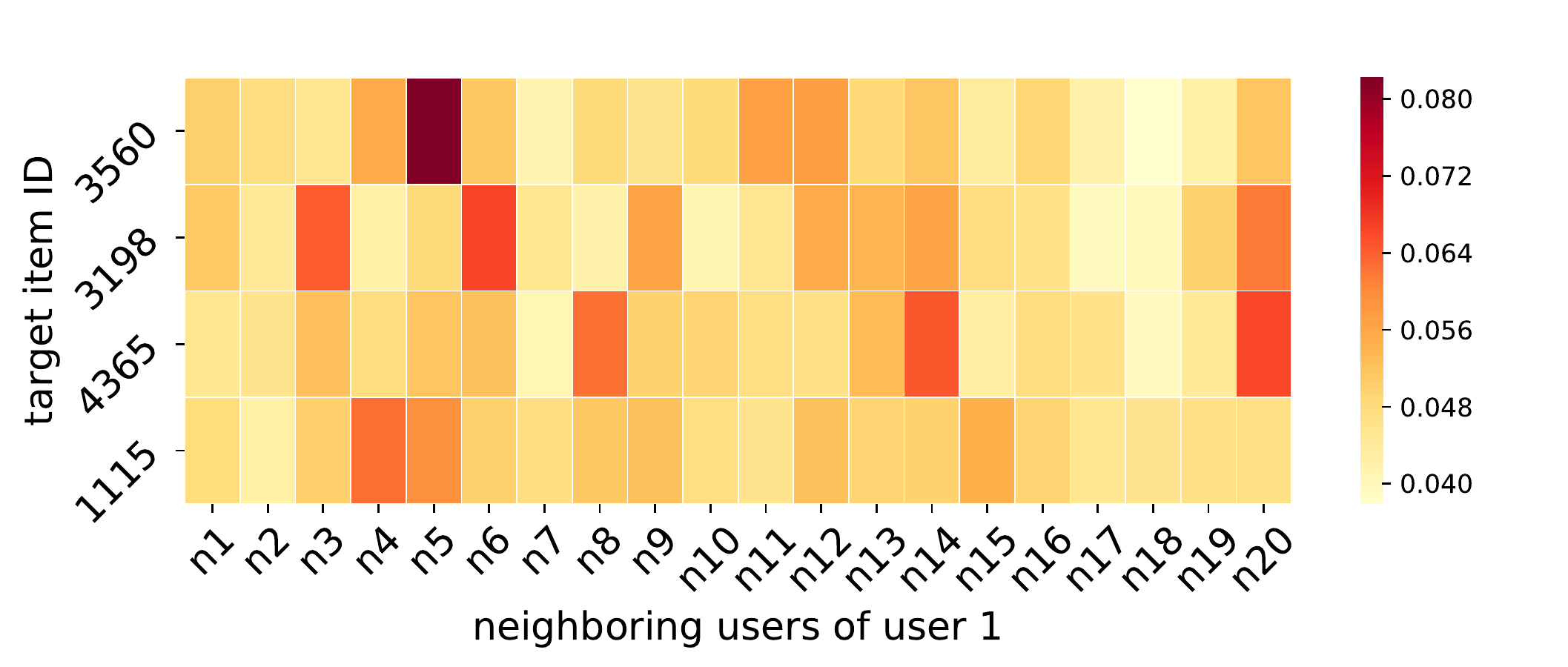}}
  \subcaptionbox{Pinterest}[.32\textwidth][c]{%
    \includegraphics[width=.32\textwidth]{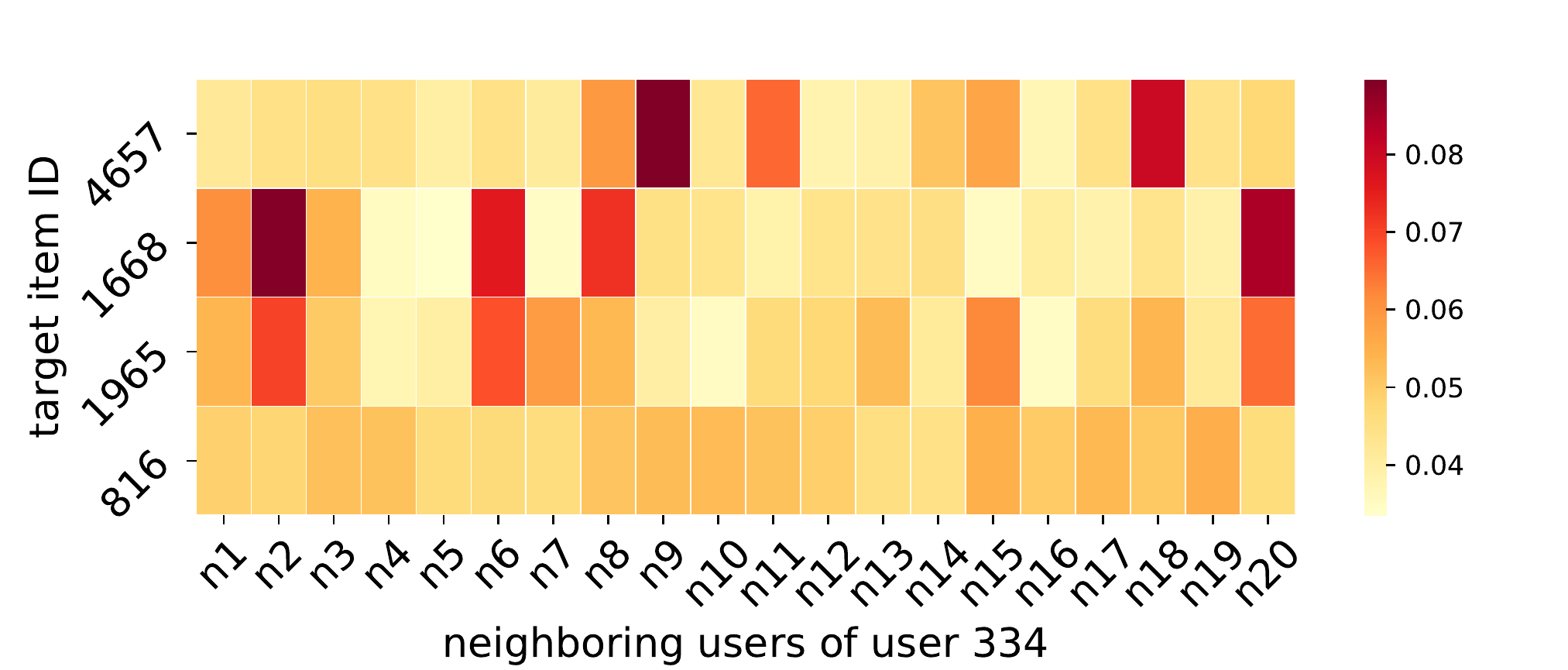}}
  \subcaptionbox{citeulike-a}[.32\textwidth][c]{%
    \includegraphics[width=.32\textwidth]{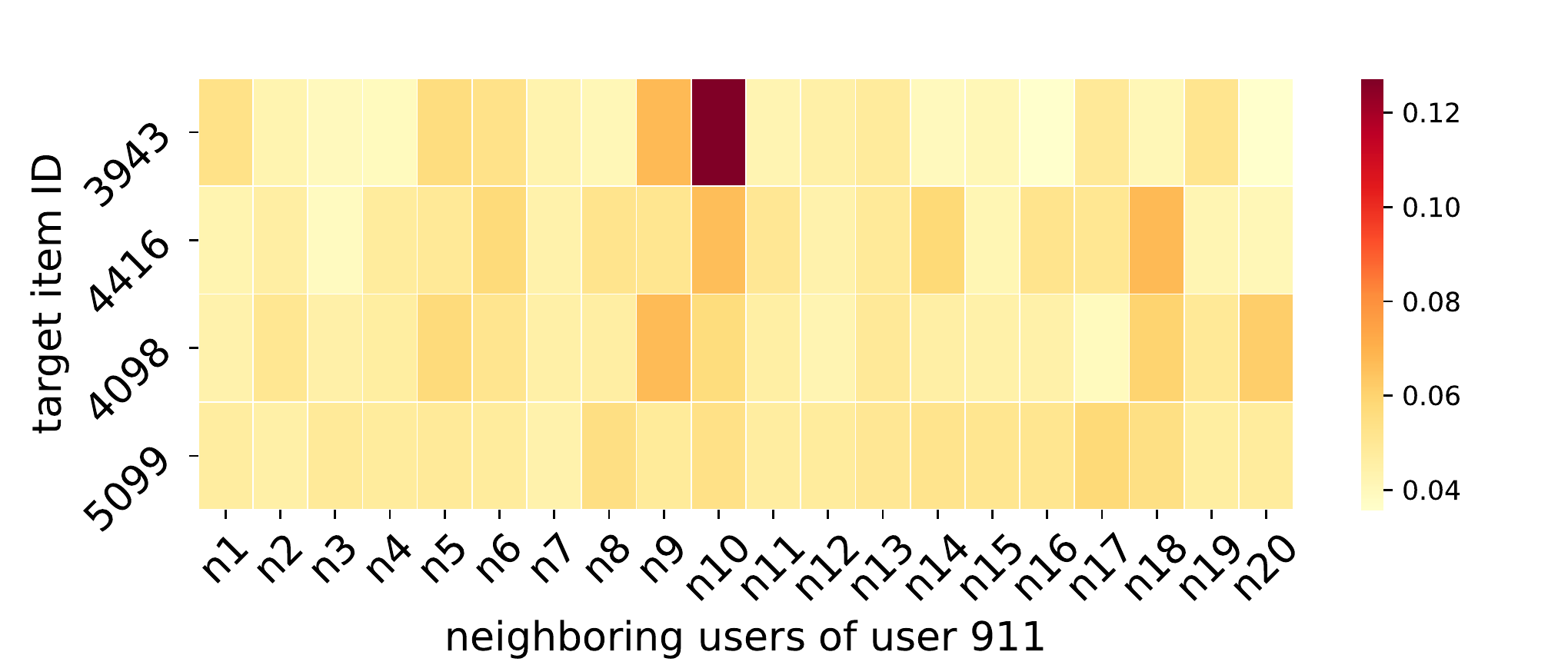}}
  \caption{Relevance scores between a randomly sampled user and his/her neighbors across the datasets, where a dark color represents a higher value while lighter colors indicate lower values. The y-axis is the target item while the x-axis illustrates the gathered neighbors given the target item. }
  \label{fig:visual}
\end{figure*}
To provide a deep insight of the attentive neighbor selection, we visualize the relevance scores (i.e. $\beta_t$ in Eqn.\ref{eq:beta}) between the users and their neighbors given different target items.  Fig.\ref{fig:visual} present a heatmap of the scores for a randomly sampled user across the datasets. The color scale indicates the intensities of the relevance scores, where a dark color represents a higher value and lighter colors indicates lower values. Each row is a score distribution over the neighbors given a target item as specified by the y-axis labels. The x-axis represents the users in the neighborhood notated from ``n1", and is truncated at ``n20". Notice that the notations do not necessarily reflect user id from the datasets.

From the figures we have the following observations. First, not all the neighbors are equally informative. For example, user 1 of the \textit{movielen-20m} dataset places a higher weight on the 5-th neighbor for estimating the ranking score of item 3560, and user 911 of the \textit{citeulike-a} depends uniquely on the 10-th neighbor to derive the recommendation of item 3943. This justifies the necessity to align the most informative neighbors for recommendation. Second, each user has different distributions of the relevance scores over the neighbors given different target items, validating the intuition of involving the target item for calculating the relevance scores. The underlying reason is that given items of different characteristics, users may refer to neighbors of different preferences to drive the recommendations. Third, in some cases, the relevance scores are evenly distributed among the neighbors, and due to the large number of neighbors, each of them receives a tiny weight for comprising the neighborhood representation. For example, the relevance scores between user 911 (\textit{citeulike-a}) and the neighbors are all close to 0 given item 5099. Similar situation can be observed between user 334 (\textit{Pinterest}) and his/her neighbors on recommending item 816. In these cases, the neighborhood information is not informative and should be automatically blocked for the recommendation task. This motivates the proposed hybrid gated network for filtering out noisy users, and select between a user representation and the neighborhood representation based on the confidence level of the neighborhood information.

\section{Related Work}
\subsection{Deep Learning in Recommendation}
Recently, deep learning has been widely applied in recommendation due to its immense success in many research areas such as computer vision, speech recognition and natural language processing \cite{lecun2015deep}. Some works \cite{nguyen2018npe} propose to boost recommendation performance by exploiting different neural network structures. He et al. \cite{he2017neural} combines generalized matrix factorization and multilayer perceptrons into a unified Neural Collaborative Filtering (NCF) framework for modeling user-item interactions. NCF is also the state-of-the-art recommendation model that is mainly based on user-item historical records. To comprehensively explore the user-item interactions, He et al. \cite{he2017nfm} propose Neural Factorization Machines (NFM) to model higher-order and non-linear feature interactions. Collaborative denoising autoencoders (CDAE) \cite{wu2016collaborative} incorporates user-specific bias into an antoencoder for recommendation, and it is proved to be able to generalize many existing collaborative filtering methods.

As those aforementioned works mainly rely on user-item interactions, they suffer from data sparseness \cite{luo2017inherently} as a user usually give few ratings compared to the large item set. To this end, many research \cite{wang2017item, ma2008sorec, guo2015trustsvd} propose to exploit additional knowledge about the users/items to mitigate data sparseness. For example, Zhang et al. \cite{zheng2017joint} employ Convolution Neural Networks (CNNs) to extract local features representations from the reviews, and utilize Factorization Machines to capture high-order interactions between representations of users and items. Wang et al. \cite{wang2017your} use pre-trained CNNs model to exploit visual content from images, and integrate those visual features for  recommending point of interest. In \cite{dong2017hybrid}, the authors leverage the effective representation learning of deep learning techniques, and propose a model that jointly performs users/items representation learning from side information and collaborative filtering from rating matrix. Meng et al. \cite{meng2018exploiting} exploit positive and negative emotions on reviews for recommendation products inspiring the fact that emotions on reviews have strong indication of user preferences. One drawback of those works is that they require side information, and it is not always effective for learning informative user/item latent factors.

\subsection{Neighborhood-based Recommendation}
Neighborhood-based approaches for recommendation is another major class of collaborative filtering. The underlying reason is that users usually show somewhat similar preferences with their neighbors, and the semantic gaps between unseen user-item pairs can be bridged by the neighbors with sufficient historical interaction records. For example, Ebesu et al. \cite{ebesu2018collaborative} proposed a recommender that is a hybrid of latent factor model and neighborhood-based model. It leverages a memory component to encode complex user-item relations, and a neural attention mechanism to learn a user-item specific neighborhood, and use a output module to jointly exploit the memory component and the neighborhood to produce the ranking score. In \cite{chen2019social}, the authors address for social-aware recommendation by using a hierarchical attention mechanism that exploits aspect-level and friend-level attentions from the neighborhood for recommendation. Ma et al. \cite{ma2019gated} address the sparse implicit feedback of recommendation by proposing a neighbor-level attention that learns the neighborhood representation of an item by considering its neighbors in a weighted manner.

Attention mechanism is an indispensable technique in those neighborhood-based approaches. It has wide application in many machine learning tasks such as image captioning and machine translation \cite{bahdanau2014neural, rush2015neural}. Since not all users/items in the neighborhood are equally informative, it is natural to place higher weights on some specific neighbors when aggregating the representations of the neighbors. However, the neighborhood is usually noisy and simply aggregating all the neighbors has negative impact on the recommendation performance. Moreover, most of the previous words tend to assign equal weights to the user and neighborhood representation, ignoring the confidence of the neighbors in recommending the target item. Finally, those previous neighborhood-based approaches ignore the semantic proximity between the users/items and their neighbors, and the user-user similarities are implicitly captured during the collaborative filtering process, and it is inefficient for modeling localized neighborhood information.
\section{Conclusion}
In this paper, we propose a novel neighborhood-based recommendation model to deal with neighborhood noises and learn compact user-neighbor compatibility. We design a hybrid gated network for separating similar neighbors from dissimilar ones, and aggregate those similar neighbors for comprising neighborhood representations. We also propose to explicitly preserve user-neighbor proximity, and learn specialized user representations for the recommendation task. Extensive experiments on three publicly datasets have demonstrated the advantage of the proposed model over state-of-the-art neighborhood-based models, and justified the rationale underlying the two components in the proposed model.
\section{Acknowledgement}
This research was supported in part by the National Natural Science Foundation of China under Grant 61802427.

\bibliographystyle{elsarticle-num}
\bibliography{bib}

\end{document}